\documentclass[aps,pra,preprint,showpacs]{revtex4}
\usepackage[dvips]{graphicx}
\def\C{{\bf C}}
\def\R{{\bf R}}
\def\Z{{\bf Z}}

\begin{document}

\title{Limit distributions of two-dimensional quantum walks}

\author{Kyohei Watabe}
%\email[]{}
\affiliation{Department of Physics,
Faculty of Science and Engineering,
Chuo University, Kasuga, Bunkyo-ku, Tokyo 112-8551, Japan}
\author{Naoki Kobayashi} 
\email[]{knaoki@phys.chuo-u.ac.jp}
\affiliation{Department of Physics,
Faculty of Science and Engineering,
Chuo University, Kasuga, Bunkyo-ku, Tokyo 112-8551, Japan}
\author{Makoto Katori} 
\email[]{katori@phys.chuo-u.ac.jp}
\affiliation{Department of Physics,
Faculty of Science and Engineering,
Chuo University, Kasuga, Bunkyo-ku, Tokyo 112-8551, Japan}
\author{Norio Konno}
\email[]{konno@ynu.ac.jp}
\affiliation{
Department of Applied Mathematics, 
Yokohama National University, 
79-5 Tokiwadai, Yokohama 240-8501, Japan}

%\date{\today}
\date{20 June 2008}

\begin{abstract}
One-parameter family of discrete-time quantum-walk models
on the square lattice, which includes
the Grover-walk model as a special case, is analytically studied.
Convergence in the long-time limit $t \rightarrow \infty$
of all joint moments of two components
of walker's pseudovelocity, 
$X_t/t$ and $Y_t/t$, is proved and the probability density
of limit distribution is derived.
Dependence of the two-dimensional limit 
density function on the parameter
of quantum coin and initial four-component qudit
of quantum walker is determined. 
Symmetry of limit distribution on a plane
and localization around the origin
are completely controlled.
Comparison with numerical results of 
direct computer-simulations is also shown.
\end{abstract}

% insert suggested PACS numbers in braces on next line
% insert suggested keywords - APS authors don't need to do this
%\keywords{}

\maketitle

%%%%%%%%%%%%%%%%%%%%%%%%%%%%%%%%%%%%%%%%%%%%%%%%%%%%%%%%%%%%
\section{INTRODUCTION}
%%%%%%%%%%%%%%%%%%%%%%%%%%%%%%%%%%%%%%%%%%%%%%%%%%%%%%%%%%%%

Quantum walks are expected to provide mathematical models
for quantum algorithms, which could be used 
in quantum computers in the future 
\cite{Tra,Kem03,Amb03,BCA03,Ken06}.
Though the systematic study of quantization of random walks
is not old \cite{ADZ93,Mey96,NV00,ABNVW01},
one-dimensional models have been well studied
and mathematical properties are clarified \cite{Kon07,Str06}.
For example, convergence of all moments of pseudovelocity
in the long-time limit was proved for
the standard two-component quantum-walk model
and the weak limit-theorem is established \cite{Kon02,Kon05,GJS04,KFK05}.
The weak limit-theorem was generalized
for the multi-component quantum-walk models
associated with rotation matrices \cite{MKK07,SKKK08}.

One of the recent topics of quantum walks is 
systematic study of higher dimensional models
\cite{MBSS02,TFMK03,GJS04,CLXGKK05,VBBB05,MVB05}.
Among them the Grover-walk model
has been extensively studied, 
since it is related to Grover's search algorithm 
\cite{Gro97,SKW03,CG04a,CG04b,Tulsi08}.
Inui $et$ $al$.\cite{IKK04} studied the two-dimensional 
Grover-walk model analytically 
and clarified an interesting phenomenon called {\it localization}
\cite{MBB07}.
In two dimensions effect of random environment 
on quantum systems is non-trivial
and {\it decoherence} in two-dimensional quantum walks 
generated by broken-line-type noise
was studied by Oliveira {\it et al.} \cite{OPD06}.

We noted that at the end of the paper by Inui $et$ $al$.\cite{IKK04} 
a one-parameter family of two-dimensional quantum-walk models 
was introduced,
which includes the Grover walk as a special case;
with the parameter $p=1/2$ of a quantum coin.
In general the quantum walker on the square lattice,
which hops to one of the four nearest-neighbor sites
at each time step,
is described by a four-component wave function.
In the present paper, we will determine
the dependence of long-time behavior of quantum walker
both on the parameter $p$ 
and a four-component initial wave function (four-component qudit)
completely and establish the weak limit-theorem 
for the family of two-dimensional models.

This paper is organized as follows.
In Sec.II we define the discrete-time 
two-dimensional quantum-walk models.
By calculating the eigenvalues and eigenvectors
of the time-evolution matrix of
quantum walk in the wave-number space,
long-time behavior of joint moments
of $x$ and $y$ components of pseudovelocity
is analyzed in Sec.III.
There the weak limit-theorem
for the two-dimensional models is proved and 
dependence of the limit distributions of pseudovelocities
on the parameter $p$ of quantum coin
and on an initial qudit of walker is clarified.
In order to demonstrate the usefulness
of our results to control
the long-time behavior of quantum walks,
we show pairs of figures of direct computer-simulation results
and of obtained limit distributions in Sec.IV.
Using the results we can discuss symmetry of limit distributions
on a plane systematically depending on the parameter $p$
and initial qudits of walker.
Concluding remarks are given in Sec.V.
Appendix A is used to show calculation of
some integrals.

%%%%%%%%%%%%%%%%%%%%%%%%%%%%%%%%%%%%%%%%%%%%%%%%%%%%%
\section{TWO-DIMENSIONAL QUANTUM WALK MODELS}
%%%%%%%%%%%%%%%%%%%%%%%%%%%%%%%%%%%%%%%%%%%%%%%%%%%%%
\subsection{General setting on the square lattice}
%%%%%%%%%%%%%%%%%%%%%%%%%%%%%%%%%%%%%%%%%%%%%%%%%%%%%
We begin with defining
the two-dimensional discrete-time quantum walk
on the square lattice $\Z^2=\{(x,y) : x,y \in \Z \}$,
where $\Z$ denotes a set of all integers
$\Z=\{\cdots, -2, -1, 0, 1, 2, \cdots\}$.
Corresponding to the fact that
there are four nearest-neighbor sites
for each site $(x,y) \in \Z^2$,
we assign a four-component wave function
$$
\Psi(x,y,t) = \left(
\begin{array}{c}
\psi_1(x,y,t) \cr
\psi_2(x,y,t) \cr
\psi_3(x,y,t) \cr
\psi_4(x,y,t) 
\end{array}
\right)
$$
to a quantum walker, each component of 
which is a complex function
of location $(x,y) \in \Z^2$ and discrete time
$t=0,1,2,\cdots$.
A quantum coin will be given by a $4 \times 4$ unitary matrix, 
$A=(A_{jk})_{j,k=1}^{4}$, 
and a spatial shift-operator on $\Z^2$
is represented in the wave-number space
$(k_x,k_y) \in [-\pi,\pi)^2$
by a matrix
\begin{eqnarray}
S(k_x,k_y) =&
\left(\begin{array}{cccc}
e^{ik_x} & 0 & 0 & 0\\
0 & e^{-ik_x} & 0 & 0\\
0 & 0 & e^{ik_y} & 0\\
0 & 0 & 0 & e^{-ik_y}
\end{array} \right),
\nonumber
\end{eqnarray}
where $i=\sqrt{-1}$. 
We assume
that at the initial time $t=0$ 
the walker is located at the origin
with a four-component qudit
${}^{\!T}\phi_0=(q_1,q_2,q_3,q_4)
\in \C^4$, $\sum_{j=1}^{4} |q_j|^2 =1$.
In the present paper, the transpose of vector/matrix
is denoted by putting a superscript $T$
on the left,
and $\R$ and $\C$ denote the sets of all real
and complex numbers, respectively.
Let
\begin{equation}
V(k_x,k_y)\equiv S(k_x,k_y) A.
\label{eqn:Vk1}
\end{equation}
Then, in the wave-number space, 
the wave function of the walker at time $t$ is given by
\begin{equation}
\hat{\Psi}(k_x,k_y,t)=\Big( V(k_x,k_y) \Big)^{t}
\phi_{0}, \quad
t=0,1,2, \cdots.
\label{eqn:Psih1}
\end{equation}
Time evolution in the real space $\Z^2$ is
then obtained by performing the Fourier transformation
\begin{eqnarray}
\Psi(x,y,t) &=& \int_{-\pi}^{\pi} \frac{dk_x}{2\pi} 
\int_{-\pi}^{\pi} \frac{dk_y}{2\pi} 
e^{i(k_x x+k_y y)} \hat{\Psi}(k_x,k_y,t). \nonumber
\end{eqnarray}
Note that the inverse Fourier transformation
should be
\begin{eqnarray}
\hat{\Psi}(k_x,k_y,t) &=& \sum_{(x,y) \in \Z^2} 
\Psi(x,y,t) e^{-i(k_x x+k_y y)}. \nonumber
\end{eqnarray}

Now the stochastic process of two-dimensional quantum walk
is defined on $\Z^2$ as follows.
Let $X_t$ and $Y_t$ be $x$ and $y$-coordinate 
of the position of the walker at time $t$, respectively.
The probability that we find the walker at site $(x,y) \in \Z^2$
at time $t$ is given by
\begin{equation}
P(x,y,t) \equiv
{\rm Prob}\Big((X_t, Y_t)=(x,y) \Big)
=\Psi^{\dagger}(x,y,t) \Psi(x,y,t),
\nonumber
\end{equation}
where $\Psi^{\dagger}(x,y,t) = {}^T\!\bar{\Psi}(x,y,t)$ is the
hermitian conjugate of $\Psi(x,y,t)$.
The joint moments of $X_t$ and $Y_t$
are given by
\begin{eqnarray}
\left\langle X_t^{\alpha} Y_t^{\beta} \right\rangle &\equiv&
\sum_{(x,y) \in \Z^2} x^{\alpha} y^{\beta} P(x,y,t)
\nonumber\\
&=& \int_{-\pi}^{\pi} \frac{dk_x}{2\pi} 
\int_{-\pi}^{\pi} \frac{dk_y}{2\pi} 
\hat{\Psi}^{\dagger}(k_x, k_y, t)
\left(i\frac{\partial}{\partial k_x} \right)^{\alpha}
\left(i\frac{\partial}{\partial k_y} \right)^{\beta}
\hat{\Psi}(k_x,k_y,t), 
\label{eqn:moment1}
\end{eqnarray}
for $\alpha,\beta=0,1,2,\cdots$.

%%%%%%%%%%%%%%%%%%%%%%%%%%%%%%%%%%%%%%%%%%%%%%%%%%%%%
\subsection{Generalized Grover walks}
%%%%%%%%%%%%%%%%%%%%%%%%%%%%%%%%%%%%%%%%%%%%%%%%%%%%%
Inui $et$ $al$.\cite{IKK04} introduced 
a one-parameter family of quantum-walk models
on $\Z^2$ as a generalization of Grover model
by specifying the quantum coin as
\begin{eqnarray}
A&=&
\left(\begin{array}{cccc}
-p & q & \sqrt{\mathstrut pq} & \sqrt{\mathstrut pq}\\
q & -p & \sqrt{\mathstrut pq} & \sqrt{\mathstrut pq}\\
\sqrt{\mathstrut pq} & \sqrt{\mathstrut pq} & -q & p\\
\sqrt{\mathstrut pq} & \sqrt{\mathstrut pq} & p & -q
\end{array} \right), \quad
q=1-p,\label{eqn:matrix1}
\end{eqnarray}
where $p \in (0,1)$.
When $p=1/2$, $A$ is reduced
to the quantum-coin matrix
used to generate the Grover walk on $\Z^2$.
In general the generator of the process (\ref{eqn:Vk1})
is given as
\begin{equation}
V(k_x,k_y) 
= \left(\begin{array}{cccc}
-p e^{ik_x} & q e^{ik_x} & \sqrt{\mathstrut pq} e^{ik_x} 
& \sqrt{\mathstrut pq} e^{ik_x}\\
q e^{-ik_x} & -p e^{-ik_x} & \sqrt{\mathstrut pq} e^{-ik_x} 
& \sqrt{\mathstrut pq} e^{-ik_x}\\
\sqrt{\mathstrut pq} e^{ik_y} & \sqrt{\mathstrut pq} e^{ik_y} 
& -q e^{ik_y} & p e^{ik_y}\\
\sqrt{\mathstrut pq} e^{-ik_y} & \sqrt{\mathstrut pq} e^{-ik_y} 
& p e^{-ik_y} & -q e^{-ik_y}
\end{array} \right),
\label{eqn:Vk2}
\end{equation}
$q=1-p, 0 < p < 1$.

%%%%%%%%%%%%%%%%%%%%%%%%%%%%%%%%%%%%%%%%%%%%%%%%%%%%%
\section{LIMIT DISTRIBUTION IN $t \rightarrow \infty$}
%%%%%%%%%%%%%%%%%%%%%%%%%%%%%%%%%%%%%%%%%%%%%%%%%%%%%
\subsection{Calculation of moments and their long-time limits}
%%%%%%%%%%%%%%%%%%%%%%%%%%%%%%%%%%%%%%%%%%%%%%%%%%%%%
In order to analyze the long-time behavior
of the present two-dimensional quantum walks,
we use the method originally given by 
Grimmett {\it et al.} \cite{GJS04},
which has been developed in \cite{KFK05,MKK07,SKKK08}.
It is easy to diagonalize the time-evolution matrix (\ref{eqn:Vk2}).
The four eigenvalues are obtained as
$$
\lambda_1 = 1, \quad
\lambda_2 = -1, \quad
\lambda_3 = e^{i\omega(k_x,k_y)}, \quad
\lambda_4 = e^{-i\omega(k_x,k_y)},
$$
where $\omega(k_x,k_y)$ is determined by the equation
\begin{equation}
\cos \omega(k_x,k_y) = -( p \cos k_x + q \cos k_y)
\label{eqn:eigenvalue1}.
\end{equation}
The eigenvectors corresponding to 
the eigenvalues $\lambda_j$, $1 \leq j \leq 4$, 
are given by the following column vectors
\begin{eqnarray}
{\bf v}_j(k_x,k_y) =N_j
\left(\begin{array}{cccc}
q(e^{ik_y} \lambda_j+1) (e^{ik_x} \lambda_j+1) (e^{-ik_y} \lambda_j+1)\\
q(e^{ik_y} \lambda_j+1) (e^{-ik_x} \lambda_j+1) (e^{-ik_y} \lambda_j+1)\\
\sqrt{\mathstrut pq}(e^{ik_y} \lambda_j+1) (e^{-ik_x} \lambda_j+1) 
(e^{ik_x} \lambda_j+1)\\
\sqrt{\mathstrut pq}(e^{-ik_x} \lambda_j+1) (e^{ik_x} \lambda_j+1) 
(e^{-ik_y} \lambda_j+1) 
\end{array}\right)
\label{eqn:eigenv}
\end{eqnarray}
with appropriate normalization factors
$N_j, 1 \leq j \leq 4$.
Define the $4 \times 4$ unitary matrix
$
R(k_x,k_y)\equiv({\bf v}_1,{\bf v}_2,{\bf v}_3,{\bf v}_4)
$
from the four column vectors (\ref{eqn:eigenv}).
Then the time-evolution matrix (\ref{eqn:Vk2})
is diagonalized, and
by the unitarity of $R(k_x,k_y)$,
$R^{\dagger}(k_x,k_y)=[R(k_x,k_y)]^{-1}$,
(\ref{eqn:Psih1}) is written as
\begin{eqnarray}
\hat{\Psi}(k_x,k_y,t)
&=&
R(k_x,k_y)
\left(\begin{array}{cccc}
\lambda_1^{t} & 0 & 0 & 0\\
0 & \lambda_2^{t} & 0 & 0\\
0 & 0 & \lambda_3^{t} & 0\\
0 & 0 & 0 & \lambda_4^{t}
\end{array}\right)
R^{\dagger}(k_x,k_y)\phi_0 \nonumber\\
&=&  \sum_{j=1}^{4} \left( \lambda_j \right)^{t} {\bf v}_j C_j(k_x,k_y),
\nonumber
\end{eqnarray}
where
$C_j(k_x,k_y) \equiv {\bf v}^{\dagger}_j(k_x,k_y)\phi_0$. 
For $\alpha,\beta=1,2,\cdots$, we see
\begin{eqnarray}
&&\left( i\frac{\partial}{\partial k_x} \right)^{\alpha} 
\left( i\frac{\partial}{\partial k_y} \right)^{\beta} 
\hat{\Psi}(k_x, k_y, t) \nonumber\\
&& \qquad =
\left( -\frac{\partial \omega(k_x,k_y)}{\partial k_x} \right)^{\alpha} 
\left( -\frac{\partial \omega(k_x,k_y)}{\partial k_y} \right)^{\beta} 
(\lambda_3)^{t} {\bf v}_3(k_x,k_y) C_3(k_x,k_y) t^{\alpha+\beta} \nonumber\\
&& \qquad +  
\left( \frac{\partial \omega(k_x,k_y)}{\partial k_x} \right)^{\alpha} 
\left( \frac{\partial \omega(k_x,k_y)}{\partial k_y} \right)^{\beta}
(\lambda_4)^t {\bf v}_4(k_x,k_y) C_4(k_x,k_y) t^{\alpha+\beta} 
+ O(t^{\alpha+\beta-1}), \nonumber
\end{eqnarray} 
since $\lambda_1=1$ and $\lambda_2=-1$ are 
independent of $k_x, k_y$.
Since $R(k_x,k_y)$ is unitary, its column vectors 
make a set of orthonormal vectors;
$
{\bf v}^{\dagger}_m(k_x,k_y) {\bf v}_{m'}(k_x,k_y)
=\delta_{m m'}.
$
Then we have
\begin{eqnarray}
&& \hat{\Psi}^{\dagger}(k_x, k_y, t)
\left( i\frac{\partial}{\partial k_x} \right)^{\alpha} 
\left( i\frac{\partial}{\partial k_y} \right)^{\beta}
\hat{\Psi}(k_x, k_y, t) \nonumber\\
&=&
\left\{ (-1)^{\alpha+\beta} |C_3(k_x,k_y)|^{2} 
+ |C_4(k_x,k_y)|^{2} \right\} 
\left( \frac{\partial \omega(k_x,k_y)}
{\partial k_x} \right)^{\alpha}
\left( \frac{\partial \omega(k_x,k_y)}
{\partial k_y} \right)^{\beta} 
t^{\alpha+\beta} +O(t^{\alpha+\beta-1}). \nonumber
\end{eqnarray}
The pseudovelocity of quantum walker at time $t$ is
defined as
\begin{equation}
{\bf V}_{t}=\left(
\frac{X_t}{t}, \frac{Y_t}{t} \right),
\quad t=1,2,3, \cdots.
\label{eqn:pseudov}
\end{equation}
Eq.(\ref{eqn:moment1}) gives
the following expression for
joint moments of $x$ and $y$ components of
pseudovelocity,
$ (X_t/t)^{\alpha} (Y_t/t)^{\beta} $, 
in the long-time limit
\begin{eqnarray}
\lim_{t \to \infty}\left\langle \left( \frac{X_t}{t} \right)^{\alpha} 
\left( \frac{Y_t}{t} \right)^{\beta} \right\rangle &=&
\int_{-\pi}^{\pi} \frac{dk_x}{2\pi} 
\int_{-\pi}^{\pi} \frac{dk_y}{2\pi}
\left\{ (-1)^{\alpha+\beta} |C_3(k_x,k_y)|^{2} + |C_4(k_x,k_y)|^{2} \right\} 
\nonumber\\
&& \quad \times
\left( \frac{\partial \omega(k_x,k_y)}{\partial k_x} \right)^{\alpha}
\left( \frac{\partial \omega(k_x,k_y)}{\partial k_y} \right)^{\beta}.
\nonumber 
\end{eqnarray}
Here from (\ref{eqn:eigenvalue1}) we have
$
\omega(k_x,k_y)= \, {\rm arccos}
\left\{ -( p \cos k_x + q \cos k_y) \right\}
$
and then
\begin{eqnarray}
\frac{\partial \omega(k_x,k_y)}{\partial k_x} &=& 
-\frac{p \sin k_x}{\sqrt{1-(p \cos k_x +q  \cos k_y)^{2}}}, \nonumber\\
\frac{\partial \omega(k_x,k_y)}{\partial k_y} &=& 
-\frac{q \sin k_y}{\sqrt{1-(p \cos k_x +q  \cos k_y)^{2}}} \nonumber
\end{eqnarray}
by the formula
$(d/dx) {\rm arccos} x = \mp 1/\sqrt{\mathstrut 1-x^2}.$ 

We change the variable of integral from $k_x,k_y$ to $v_x,v_y$ by
\begin{eqnarray}
v_x &=& \frac{p \sin k_x}{\sqrt{1-(p \cos k_x +q  \cos k_y)^{2}}},
\nonumber\\
v_y &=& \frac{q \sin k_y}{\sqrt{1-(p \cos k_x +q  \cos k_y)^{2}}} 
\label{eqn:change}.
\end{eqnarray}
It should be noted that this map
$(k_x, k_y) \in [-\pi, \pi)^2 \mapsto (v_x, v_y)$ is 
one-to-two and the image is a union of interior points of
an ellipse
\begin{equation}
\frac{v_x^2}{p}+\frac{v_y^2}{q} < 1
\label{eqn:ellipse}
\end{equation}
and the four points
$\{(p,q), (p, -q), (-p, q), (-p, -q)\}$.
We found that the following relations
are derived from (\ref{eqn:change}),
\begin{eqnarray}
\sin k_x &=& \frac{2v_x \sqrt{pq-qv_x^{2}-pv_y^{2}}}
{p\sqrt{(v_x+v_y+1)(v_x-v_y+1)(v_x+v_y-1)(v_x-v_y-1)}}, \nonumber\\
\cos k_x &=& \frac{(1+q)v_x^{2}+pv_y^{2}-p}
{p\sqrt{(v_x+v_y+1)(v_x-v_y+1)(v_x+v_y-1)(v_x-v_y-1)}}, \nonumber\\
\sin k_y &=& \frac{2v_y \sqrt{pq-qv_x^{2}-pv_y^{2}}}
{q\sqrt{(v_x+v_y+1)(v_x-v_y+1)(v_x+v_y-1)(v_x-v_y-1)}}, \nonumber\\
\cos k_y &=& -\frac{qv_x^{2}+(1+p)v_y^{2}-q}
{q\sqrt{(v_x+v_y+1)(v_x-v_y+1)(v_x+v_y-1)(v_x-v_y-1)}}.
\label{eqn:mapA}
\end{eqnarray}
They are useful to calculate the Jacobian
associated with the inverse map
$(v_x,v_y) \mapsto (k_x,k_y) $
and we have obtained
\begin{eqnarray}
J &\equiv&
\left|\begin{array}{rr}
\partial v_x/\partial k_x & \partial v_x/\partial k_y\\
\partial v_y/\partial k_x & \partial v_y/\partial k_y
\end{array}\right|
\nonumber\\
&=& \frac{1}{4}
\Big| (v_x+v_y+1)(v_x-v_y+1)(v_x+v_y-1)(v_x-v_y-1) \Big|.
\nonumber
\end{eqnarray}
If we assume that by this change of variable
$C_j(k_x,k_y)$ are replaced by $\hat{C}_j(v_x,v_y)$, $j=3,4$,
the integral is written as
\begin{eqnarray}
&&\lim_{t \to \infty}\left\langle \left( \frac{X_t}{t} \right)^{\alpha} 
\left( \frac{Y_t}{t} \right)^{\beta} \right\rangle \nonumber\\
&=&2 \int_{-\infty}^{\infty} \frac{dv_x}{2\pi} 
\int_{-\infty}^{\infty} \frac{dv_y}{2\pi} 
\frac{1}{J} \left\{ |\hat{C}_3(v_x,v_y)|^{2} 
+(-1)^{\alpha+\beta}|\hat{C}_4(v_x,v_y)|^{2} \right\} 
v_x^{\alpha} v_y^{\beta}
{\bf 1}_{\{v_x^2/p+v_y^2/q < 1\}}
\nonumber\\
&=&\int_{-\infty}^{\infty} dv_x 
\int_{-\infty}^{\infty} dv_y 
v_x^{\alpha} v_y^{\beta}
\mu_p(v_x,v_y){\cal M}(v_x,v_y),
\label{eqn:limit1}
\end{eqnarray}
where ${\bf 1}_{\{\Omega\}}$ denotes the
indicator function of a condition $\Omega$;
${\bf 1}_{\{\Omega\}}=1$ if $\Omega$ is satisfied
and ${\bf 1}_{\{\Omega\}}=0$ otherwise.
Here $\mu_p(v_x,v_y)$ is given by
\begin{eqnarray}
\mu_p(v_x,v_y)=\frac{2}{\pi^{2} 
(v_x+v_y+1)(v_x-v_y+1)(v_x+v_y-1)(v_x-v_y-1)} 
{\bf 1}_{\{v_x^{2}/p+v_y^{2}/q < 1 \} },
\label{eqn:bunpu1}
\end{eqnarray}
since we can confirm that
$(v_x+v_y+1)(v_x-v_y+1)(v_x+v_y-1)(v_x-v_y-1)>0$, 
when $v_x^2/p+v_y^2/q<1, q=1-p, 0 < p < 1$.
This function $\mu_p(v_x, v_y)$ gives the fundamental
density-function for long-time limit distribution
of pseudovelocity (see Appendix A).
Figure \ref{fig:fig_mup} shows it when $p=1/4$.
It should be noted that the fundamental density-function
$\mu_p(v_x, v_y)$ depends on the parameter $p$
but does not on an initial qudit
${}^{\!T}\phi_0=(q_1, q_2, q_3, q_4)$.
The dependence on an initial qudit is expressed by
the weight function ${\cal M}(v_x, v_y)$ given below.
%%%%%%%%%%%%%%%%%%%%%%%%%%%%%%%%%%%%%%%%%%%%%%%%%%%%%%%%
\begin{figure}[htpb]
\begin{center}
\includegraphics[width=1.0\linewidth]{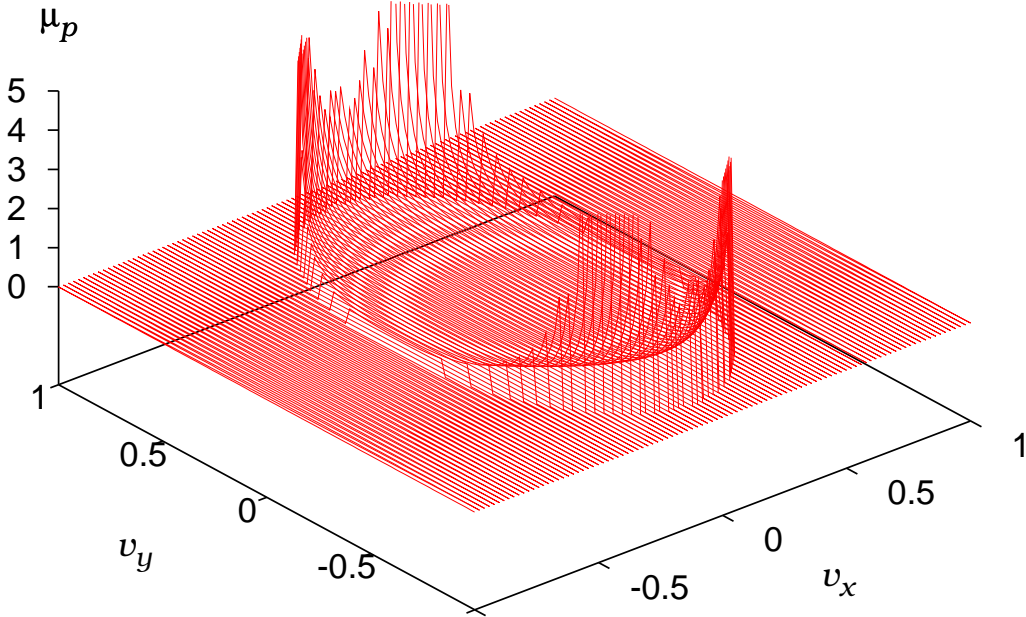}
\caption{(Color online) 
The two-dimensional fundamental density-function
$\mu_p(v_x, v_y)$ 
of limit distribution of pseudovelocities, when $p=1/4$.}
\label{fig:fig_mup}
\end{center}
\end{figure}
%%%%%%%%%%%%%%%%%%%%%%%%%%%%%%%%%%%%%%%%%%%%%%%%%%%%%%%%

%%%%%%%%%%%%%%%%%%%%%%%%%%%%%%%%%%%%%%%%%%%%%%%%%%%%%%%%%%%%%%%%%%%%%
\subsection{Weight function ${\cal M}(v_x, v_y)$}
%%%%%%%%%%%%%%%%%%%%%%%%%%%%%%%%%%%%%%%%%%%%%%%%%%%%%%%%%%%%%%%%%%%%%

Using (\ref{eqn:mapA}), the weight function
${\cal M}(v_x,v_y)$
is explicitly determined as follows;
\begin{equation}
{\cal M}(v_x,v_y)={\cal M}_1+{\cal M}_2 v_x
+{\cal M}_3 v_y +{\cal M}_4 v_x^{2}+{\cal M}_5 v_y^{2}
+{\cal M}_6 v_x v_y
\label{eqn:poly1}
\end{equation}
with
\begin{eqnarray}
\mathcal{M}_1 &=& \frac{1}{2} +
\mathrm{Re}(q_1\bar{q}_2+q_3\bar{q}_4),\nonumber\\
\mathcal{M}_2 &=& -\Big( |q_1|^{2}-|q_2|^{2} \Big) 
+ \frac{q}{\sqrt{\mathstrut pq}} 
\mathrm{Re}(q_1\bar{q}_3+q_1\bar{q}_4-q_2\bar{q}_3-q_2\bar{q}_4),
\nonumber\\
\mathcal{M}_3 &=& -\Big( |q_3|^{2}-|q_4|^{2} \Big) 
+ \frac{p}{\sqrt{\mathstrut pq}} 
\mathrm{Re}(q_1\bar{q}_3-q_1\bar{q}_4+q_2\bar{q}_3-q_2\bar{q}_4),
\nonumber\\
\mathcal{M}_4 &=&\frac{1}{2} 
\Big( |q_1|^{2}+|q_2|^{2}-|q_3|^{2}-|q_4|^{2} \Big) 
- \frac{1+q}{p} \mathrm{Re}(q_1\bar{q}_2) 
-\mathrm{Re}(q_3\bar{q}_4)\nonumber\\
&-&\frac{q}{\sqrt{\mathstrut pq}} 
\mathrm{Re}(q_1\bar{q}_3+q_1\bar{q}_4+q_2\bar{q}_3+q_2\bar{q}_4),
\nonumber\\
\mathcal{M}_5 &=&-\frac{1}{2} 
\Big( |q_1|^{2}+|q_2|^{2}-|q_3|^{2}-|q_4|^{2} \Big) 
- \mathrm{Re}(q_1\bar{q}_2) -\frac{1+p}{q}\mathrm{Re}(q_3\bar{q}_4)
\nonumber\\
&-&\frac{p}{\sqrt{\mathstrut pq}} 
\mathrm{Re}(q_1\bar{q}_3+q_1\bar{q}_4+q_2\bar{q}_3+q_2\bar{q}_4),
\nonumber\\
\mathcal{M}_6 &=&-\frac{1}{\sqrt{\mathstrut pq}} 
\mathrm{Re}(q_1\bar{q}_3-q_1\bar{q}_4-q_2\bar{q}_3+q_2\bar{q}_4),
\label{eqn:poly2}
\end{eqnarray}
where ${\rm Re}(z)$ denotes the real part of $z \in \C$ 
and $\overline{z}$ denotes the complex conjugate
of $z \in \C$.
The weight function defines the following 
real symmetric matrices ${\bf{M}}_n$,
through the relations
$\mathcal{M}_n=\phi^{\dagger}_0 {\bf{M}}_n \phi_0$,
$1 \leq n \leq 6$, 
\begin{eqnarray}
{\bf{M}}_1 
&=&\frac{1}{2}
\left(\begin{array}{cccc}
1 & 1 & 0 & 0\\
1 & 1 & 0 & 0\\
0 & 0 & 1 & 1\\
0 & 0 & 1 & 1
\end{array}\right),\quad
{\bf{M}}_2
=-\frac{1}{2\sqrt{\mathstrut pq}}
\left(\begin{array}{cccc}
2\sqrt{\mathstrut pq} & 0 & -q & -q\\
0 & -2\sqrt{\mathstrut pq} & q & q\\
-q & q & 0 & 0\\
-q & q & 0 & 0
\end{array}\right),\nonumber\\
{\bf{M}}_3
&=&-\frac{1}{2\sqrt{\mathstrut pq}}
\left(\begin{array}{cccc}
0 & 0 & -p & p\\
0 & 0 & -p & p\\
-p & -p & 2\sqrt{\mathstrut pq} & 0\\
p & p & 0 & -2\sqrt{\mathstrut pq}
\end{array}\right),\quad
{\bf{M}}_4
=-\frac{1}{2}
\left(\begin{array}{cccc}
-1 & \frac{1+q}{p} & \frac{q}{\sqrt{pq}} & \frac{q}{\sqrt{pq}}\\
\frac{1+q}{p} & -1 & \frac{q}{\sqrt{pq}} & \frac{q}{\sqrt{pq}}\\
\frac{q}{\sqrt{pq}} & \frac{q}{\sqrt{pq}} & 1 & 1\\
\frac{q}{\sqrt{pq}} & \frac{q}{\sqrt{pq}} & 1 & 1
\end{array}\right),\nonumber\\
{\bf{M}}_5
&=&-\frac{1}{2}
\left(\begin{array}{cccc}
1 & 1 & \frac{p}{\sqrt{pq}} & \frac{p}{\sqrt{pq}}\\
1 & 1 & \frac{p}{\sqrt{pq}} & \frac{p}{\sqrt{pq}}\\
\frac{p}{\sqrt{pq}} & \frac{p}{\sqrt{pq}} & -1 & \frac{1+p}{q}\\
\frac{p}{\sqrt{pq}} & \frac{p}{\sqrt{pq}} & \frac{1+p}{q} & -1
\end{array}\right),\quad
{\bf{M}}_6
=\frac{1}{2\sqrt{\mathstrut pq}}
\left(\begin{array}{cccc}
0 & 0 & -1 & 1\\
0 & 0 & 1 & -1\\
-1 & 1 & 0 & 0\\
1 & -1 & 0 & 0
\end{array}\right).\nonumber
\end{eqnarray}
Such matrix representations will be useful,
when we generalize the present results
to other models,
whose quantum coins are given by
larger matrices \cite{SKKK08}.

The integral
$\int_{-\infty}^{\infty} dv_x 
\int_{-\infty}^{\infty} dv_y 
\mu_p(v_x,v_y)  {\cal M}(v_x,v_y)$
is generally less than one,
since the contributions from the eigenvalues $\lambda_1$ and $\lambda_2$
have not been included.
The difference
\begin{equation}
\Delta =1-\int_{-\infty}^{\infty} dv_x 
\int_{-\infty}^{\infty} dv_y
\mu_p(v_x,v_y)  {\cal M}(v_x,v_y)
\label{eqn:Delta0}
\end{equation}
gives the weight of a point mass at $v_x=v_y=0$ in the distribution.
That is, $\Delta$ gives the probability of 
localization around the origin of the present two-dimensional
quantum walks \cite{IKK04,MKK07}
(see Sec.III.D below).

%%%%%%%%%%%%%%%%%%%%%%%%%%%%%%%%%%%%%%%%%%%%%%%%%%%%%%%%%%%%%%%%%%%%%
\subsection{Weak limit-theorem and symmetry of limit distribution}
%%%%%%%%%%%%%%%%%%%%%%%%%%%%%%%%%%%%%%%%%%%%%%%%%%%%%%%%%%%%%%%%%%%%%

The result is summarized as the following limit theorem.

\noindent{\bf Theorem} \quad
Let 
\begin{equation}
\nu(v_x,v_y)=\mu_p(v_x,v_y)
{\cal M}(v_x,v_y)+\Delta \delta(v_x) \delta(v_y),
\label{eqn:nu1}
\end{equation}
where $\mu_p(v_x,v_y)$, ${\cal M}(v_x,v_y)$,
and $\Delta$ are
given by (\ref{eqn:bunpu1}), 
(\ref{eqn:poly1}) with (\ref{eqn:poly2}),
and (\ref{eqn:Delta0}), respectively,
and $\delta(z)$ denotes
Dirac's delta function.
Then
\begin{equation}
\lim_{t \to \infty} \left\langle 
\left( \frac{X_t}{t} \right)^{\alpha} 
\left( \frac{Y_t}{t} \right)^{\beta} \right\rangle =
\int_{-\infty}^{\infty} 
dv_x \int_{-\infty}^{\infty} dv_y 
v_x^{\alpha} v_y^{\beta}
\nu(v_x,v_y)
\label{eqn:limit2}
\end{equation}
for all $\alpha, \beta=0,1,2,\cdots$.

As mentioned in an earlier paper \cite{MKK07},
distribution of quantum walks itself does not
converge in the long-time limit, since 
time evolution of quantum system is simply given by
a unitary transformation.
The above theorem is regarded as 
a {\it weak} limit-theorem in the sense that,
if we evaluate {\it moments} of pseudovelocity
in oscillatory distributions
of realized quantum walks,
the results shall be converge to the values
calculated by the formula (\ref{eqn:limit2})
with the density function (\ref{eqn:nu1})
in $t \to \infty$.
If we integrate $\nu(v_x, v_y)$ over any
region $D$ on a plane $\R^2$,
then we obtain the probability that
the pseudovelocity 
${\bf V}_t=(X_t/t, Y_t/t) \in D$
in the $t \to \infty$ limit.

The polynomial form of (\ref{eqn:poly1})
leads to
the following classification of symmetry realized in
the limit distribution.
\begin{description}
\item{(i)} \quad
When $\mathcal{M}_3=\mathcal{M}_6=0$,
the limit of probability density $\nu(v_x,v_y)$
has the reflection symmetry for the $v_x$-axis;
$\nu(v_x,-v_y)=\nu(v_x,v_y)$.

\item{(ii)} \quad
When $\mathcal{M}_2=\mathcal{M}_6=0$,
the limit of probability density $\nu(v_x,v_y)$
has the reflection symmetry for the $v_y$-axis;
$\nu(-v_x,v_y)=\nu(v_x,v_y)$.

\item{(iii)} \quad
When $\mathcal{M}_2=\mathcal{M}_3=\mathcal{M}_6=0$,
the limit of probability density $\nu(v_x,v_y)$
has the reflection symmetries 
both for the $v_x$-axis and the $v_y$-axis;
$\nu(v_x,-v_y)=\nu(-v_x,v_y)=\nu(v_x,v_y)$.

\item{(iv)} \quad
When $\mathcal{M}_2=\mathcal{M}_3=0$,
the limit of probability density $\nu(v_x,v_y)$
has the bi-rotational symmetry for the $v_z$-axis,
which is perpendicular both to $v_x$- and $v_y$-axes;
$\nu(-v_x,-v_y)=\nu(v_x,v_y)$.
\end{description}

%%%%%%%%%%%%%%%%%%%%%%%%%%%%%%%%%%%%%%%%%%%%%%%%%%%%%%%%%%%%%%%%%%%%%
\subsection{Localization probability around the origin}
%%%%%%%%%%%%%%%%%%%%%%%%%%%%%%%%%%%%%%%%%%%%%%%%%%%%%%%%%%%%%%%%%%%%%

By symmetry of the fundamental density-function
(\ref{eqn:bunpu1}),
(\ref{eqn:Delta0}) with (\ref{eqn:poly1}) becomes
$$
\Delta=1-{\cal M}_1 -{\cal M}_4 K_x
-{\cal M}_5 K_y
$$
with
\begin{eqnarray}
K_x &=& \int_{-\infty}^{\infty} dv_x
\int_{-\infty}^{\infty} dv_y
\mu_p(v_x, v_y) v_x^2,
\nonumber\\
K_y &=& \int_{-\infty}^{\infty} dv_x
\int_{-\infty}^{\infty} dv_y
\mu_p(v_x, v_y) v_y^2.
\nonumber
\end{eqnarray}
As shown in Appendix A, these integrals are
readily performed and we obtain the
following explicit expression
for the probability of localization around the origin,
\begin{equation}
\Delta = 1- {\cal M}_1 
-\frac{2}{\pi} (\arcsin \sqrt{p}-\sqrt{pq})
{\cal M}_4
-\frac{2}{\pi} (\arcsin \sqrt{q}-\sqrt{pq})
{\cal M}_5.
\label{eqn:Delta1}
\end{equation}

\begin{figure}[htpb]
\begin{center}
\includegraphics[width=1.0\linewidth]{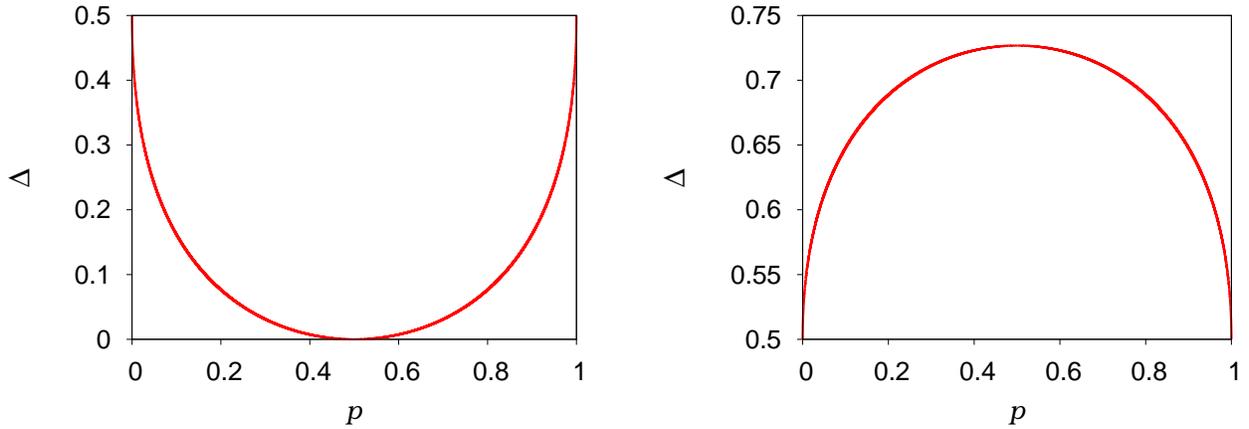}
\caption{(Color online)
Dependence of localization probability 
around the origin $\Delta$ on the parameter
$p \in (0,1)$.
(a) The case ${}^{T}\!\phi_0=(1,1,-1,-1)/2$.
When $p=1/2$ (the Grover-walk model),
$\Delta=0$.
(b) The case ${}^{T}\!\phi_0=(1,1,1,1)/2$.
When $p=1/2$ (the Grover-walk model),
$\Delta=2(\pi-2)/\pi=0.726 \cdots$.
\label{fig:Fig_Delta}}
\end{center}
\end{figure}

The localization probability $\Delta$ is a function
of the parameter $p \in (0,1)$
and an initial four-component qudit
${}^{T}\!\phi_0
=(q_1, q_2, q_3, q_4) \in \C^4,
\sum_{j=1}^{4}|q_j|^2=1$
through (\ref{eqn:poly2}).
For example, (\ref{eqn:Delta1}) gives
$$
\Delta=\frac{1}{\pi}(1-2\sqrt{pq})
\left\{ \frac{1}{p} \arcsin \sqrt{p}
+\frac{1}{q} \arcsin \sqrt{q}
-\frac{1}{\sqrt{pq}} \right\}
$$
for ${}^{T}\!\phi_0=(1,1,-1,-1)/2$,
and
$$
\Delta=\frac{1}{\pi}(1+2\sqrt{pq})
\left\{ \frac{1}{p} \arcsin \sqrt{p}
+\frac{1}{q} \arcsin \sqrt{q}
-\frac{1}{\sqrt{pq}} \right\}
$$
for ${}^{T}\!\phi_0=(1,1,1,1)/2$,
respectively, where $q=1-p$.
As shown in Fig.\ref{fig:Fig_Delta}, 
for ${}^{T}\!\phi_0=(1,1,-1,-1)/2$,
the localization probability $\Delta$
attains the minimum $=0$
for the Grover-walk model, $p=q=1/2$,
while for ${}^{T}\!\phi_0=(1,1,1,1)/2$,
it attains the maximum $=2(\pi-2)/\pi=0.726 \cdots$
for the Grover-walk model.

%%%%%%%%%%%%%%%%%%%%%%%%%%%%%%%%%%%%%%%%%%%%%%%%%%%%%%%%
\vskip 0.5cm
%%%%%%%%%%%%%%%%%%%%%%%%%%%%%%%%%%%%%%%%%%%%%%%%%%%%%%%%
%%%%%%%%%%%%%%%%%%%%%%%%%%%%%%%%%%%%%%%%%%%%%%%%%%%%%%%%
\vskip 0.5cm
%%%%%%%%%%%%%%%%%%%%%%%%%%%%%%%%%%%%%%%%%%%%%%%%%%%%%%%%

If we make the initial qudit depend on the 
parameter as
\begin{equation}
{}^{T}\!\phi_0
=\left(\sqrt{\frac{p}{2}}, \sqrt{\frac{p}{2}},
-\sqrt{\frac{q}{2}}, -\sqrt{\frac{q}{2}} \right),
\quad q=1-p,
\label{eqn:specialI}
\end{equation}
for example, then $\Delta \equiv 0$ for
${\cal M}_1=1, {\cal M}_4={\cal M}_5=0$,
and thus the quantum walker is
{\it extended} with probability one
for all $p \in (0,1)$.

It should be noted that $\Delta$ is defined
as the intensity of Dirac's delta-function at the origin
found in the limit density-function of
pseudovelocity (see Eq.(\ref{eqn:nu1})).
It implies that $\Delta$ gives the probability that
the quantum walker loses its velocity
and stays {\it around} the starting point,
{\it i.e.} the origin.
Therefore, $\Delta$ is, in general, greater than 
the time-averaged probability that the walker
stays exactly at the starting point,
$\overline{P}_{\infty}$, 
which was calculated in \cite{IKK04}.
For example, for the Grover-walk model
with the initial qudit
${}^{T}\!\phi_0=(1,1,1,1)/2$,
$\Delta=2(\pi-2)/\pi=0.726 \cdots$,
as mentioned above,
while $\overline{P}_{\infty}=2\{(\pi-2)/\pi\}^2
=0.264...$ as reported in
Sec.V.C in \cite{IKK04}.

%%%%%%%%%%%%%%%%%%%%%%%%%%%%%%%%%%%%%%%%%%%%%%%%%%%%%%%%%%%%
\section{COMPARISON WITH COMPUTER SIMULATIONS}
%%%%%%%%%%%%%%%%%%%%%%%%%%%%%%%%%%%%%%%%%%%%%%%%%%%%%%%%%%%%

\begin{figure}[htpb]
\begin{center}
\includegraphics[width=1.0\linewidth]{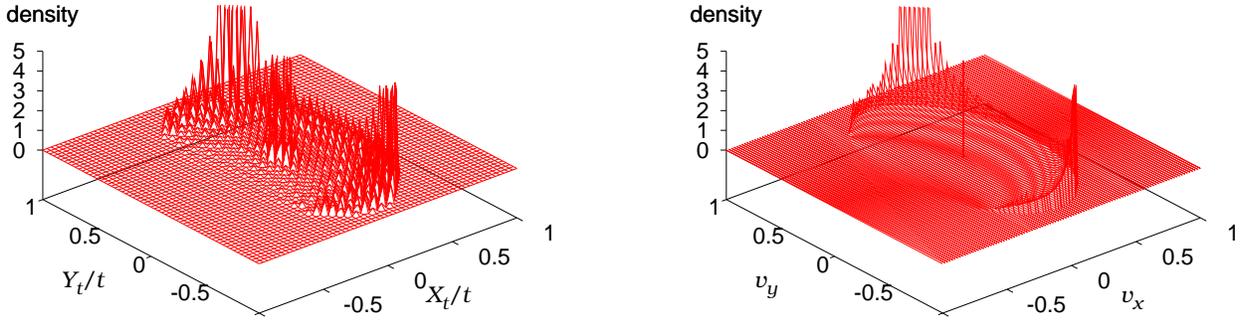}
\caption{(Color online)
The case $p=1/4$ and ${}^T\!\phi_0=(1,-1,1,1)/2$.
Since $\mathcal{M}_3=\mathcal{M}_6=0$ in this case,
the limit distribution has
the reflection symmetry for the $v_x$-axis;
$\nu(v_x,-v_y)=\nu(v_x,v_y)$.
(a) Distribution of 
pseudovelocity ${\bf V}_t=(X_t/t, Y_t/t)$ at time step $t=30$
numerically obtained by computer simulation.
(b) Probability density of 
limit distribution.
\label{fig:Fig_s_x}}
\end{center}
\end{figure}

\begin{figure}[htpb]
\begin{center}
\includegraphics[width=1.0\linewidth]{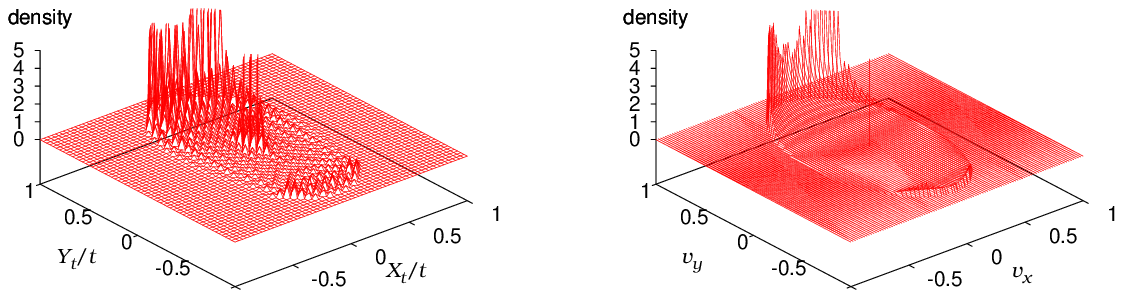}
\caption{(Color online)
The case $p=1/4$ and ${}^T\!\phi_0=(1,1,1,-1)/2$.
Since $\mathcal{M}_2=\mathcal{M}_6=0$ in this case,
the limit distribution has
the reflection symmetry for the $v_y$-axis;
$\nu(-v_x,v_y)=\nu(v_x,v_y)$.
(a) Distribution of 
pseudovelocity ${\bf V}_t=(X_t/t, Y_t/t)$
at time step $t=30$
numerically obtained by computer simulation.
(b) Probability density of 
limit distribution.
\label{fig:Fig_s_y}}
\end{center}
\end{figure}

\begin{figure}[htpb]
\begin{center}
\includegraphics[width=1.0\linewidth]{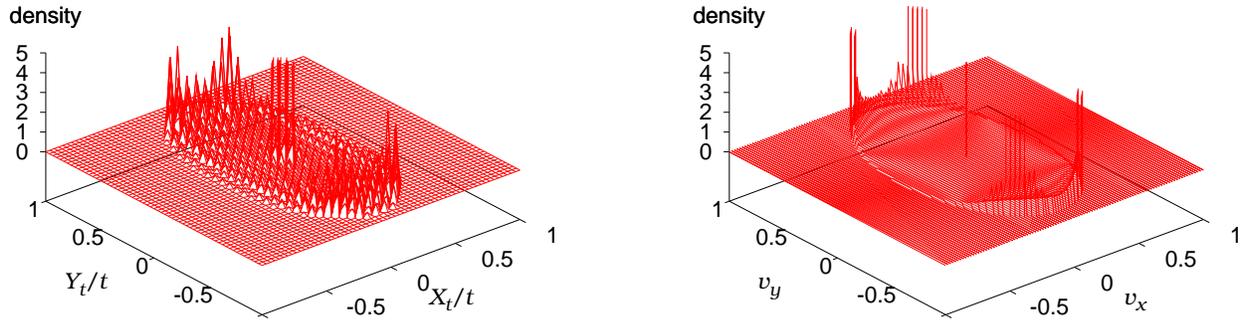}
\caption{(Color online)
The case $p=1/4$ and ${}^T\!\phi_0=(1,1,0,0)/\sqrt{2}$.
Since $\mathcal{M}_2=\mathcal{M}_3=\mathcal{M}_6=0$ in this case,
the limit distribution has
the reflection symmetries
both for the $v_x$-axis and the $v_y$-axis;
$\nu(v_x,-v_y)=\nu(-v_x,v_y)=\nu(v_x,v_y)$.
(a) Distribution of 
pseudovelocity ${\bf V}_t=(X_t/t, Y_t/t)$
at time step $t=30$
numerically obtained by computer simulation.
(b) Probability density of 
limit distribution.
\label{fig:Fig_s_xy}}
\end{center}
\end{figure}

\begin{figure}[htpb]
\begin{center}
\includegraphics[width=1.0\linewidth]{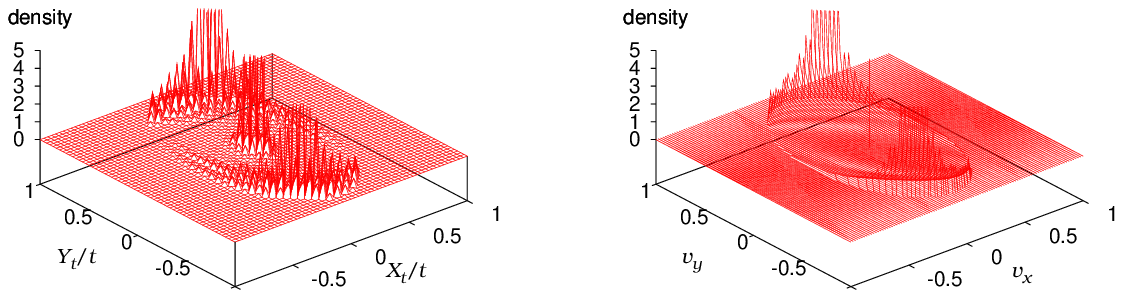}
\caption{(Color online)
The case $p=1/4$ and ${}^T\!\phi_0=(1,-1,-1,1)/2$.
Since $\mathcal{M}_2=\mathcal{M}_3=0$ in this case,
the limit distribution has
the bi-rotational symmetry for the $v_z$-axis,
which is perpendicular both to $v_x$- and $v_y$-axes;
$\nu(-v_x,-v_y)=\nu(v_x,v_y)$
(a) Distribution of 
pseudovelocity ${\bf V}_t=(X_t/t, Y_t/t)$
at time step $t=30$
numerically obtained by computer simulation.
(b) Probability density of 
limit distribution.
\label{fig:Fig_rot}}
\end{center}
\end{figure}

In order to demonstrate the validity of the above results,
here we show comparison with numerical results of
direct computer simulations \cite{MKK07}.
In Figs.\ref{fig:Fig_s_x}-\ref{fig:Fig_rot},
the left figures show
the distribution of 
pseudovelocity
${\bf V}_t=(X_t/t, Y_t/t)$ at time step $t=30$
numerically obtained by computer simulations
and the right figures the long-time limits of probability densities 
$\nu(v_x,v_y)$ determined by our theorem.
The four figures show the symmetries 
(i)-(iv) classified in Sec.III.C.
In all of these four cases shown in 
Figs.\ref{fig:Fig_s_x}-\ref{fig:Fig_rot},
$\Delta >0$ and we can see a peak at the origin
in each right figure (b), which indicates the
contribution $\Delta \delta(v_x) \delta(v_y)$
in the limit density-function (\ref{eqn:nu1}).

\vskip 0.5cm
%%%%%%%%%%%%%%%%%%%%%%%%%%%%%%%%%%%%%%%%%%%%%%%%%%%%%%%%
%%%%%%%%%%%%%%%%%%%%%%%%%%%%%%%%%%%%%%%%%%%%%%%%%%%%%%%%
\vskip 0.5cm
%%%%%%%%%%%%%%%%%%%%%%%%%%%%%%%%%%%%%%%%%%%%%%%%%%%%%%%%
%%%%%%%%%%%%%%%%%%%%%%%%%%%%%%%%%%%%%%%%%%%%%%%%%%%%%%%%
\vskip 0.5cm
%%%%%%%%%%%%%%%%%%%%%%%%%%%%%%%%%%%%%%%%%%%%%%%%%%%%%%%%
%%%%%%%%%%%%%%%%%%%%%%%%%%%%%%%%%%%%%%%%%%%%%%%%%%%%%%%%
\vskip 0.5cm
%%%%%%%%%%%%%%%%%%%%%%%%%%%%%%%%%%%%%%%%%%%%%%%%%%%%%%%%
%%%%%%%%%%%%%%%%%%%%%%%%%%%%%%%%%%%%%%%%%%%%%%%%%%%%%%%%

We observe oscillatory behavior in distributions
of ${\bf V}_t=(X_{t}/t, Y_t/t)$ in computer simulations.
In general, as the time step $t$ increases,
the frequency of oscillation becomes higher,
but, if we smear out
the oscillatory behavior, the averaged values
of distribution shall be
well-described by the density functions of limit distributions
(\ref{eqn:nu1}),
which is the phenomenon implied by our weak limit-theorem
\cite{MKK07}.

%%%%%%%%%%%%%%%%%%%%%%%%%%%%%%%%%%%%%%%%%%%%%%%%%%%%%
\section{CONCLUDING REMARKS}
%%%%%%%%%%%%%%%%%%%%%%%%%%%%%%%%%%%%%%%%%%%%%%%%%%%%%

In general, quantum coins, which determine time-evolution
of quantum walkers with spatial shift-operators,
are given by unitary transformations
\cite{BCA03,MKK07}.
The set of all $N \times N$ unitary matrices makes 
a group, the unitary group U($N$), whose dimension
is $N^2$ (see, for example, \cite{Geo99}).
Though the determinant of unitary matrix is
generally given by $e^{i \varphi},
\varphi \in [-\pi/2, \pi/2)$, this global phase factor
of quantum coin is irrelevant in
calculating any moments of walker's positions
in quantum-walk models \cite{KFK05}.
For example, in the standard two-component ($N=2$)
quantum walks, the number of relevant parameters
to specify a quantum coin is 
$N^2-1=2^2-1=3$ (Cayley-Klein parameters),
and the dependence of limit distributions of
pseudovelocities on the three parameters
was completely 
determined \cite{Kon02,Kon05,Kon07,KFK05,MKK07}.
In the present paper we have considered a one-parameter
family of unitary matrices (\ref{eqn:matrix1})
in U(4) as quantum coins. The present study should be extended
to more general models, whose U(4)-quantum coins
are fully controlled by $4^2-1=15$ 
parameters.

One of the motivations to study the present
family of models in this paper is the fact that
it contains the Grover walk on the plane.
It will be interesting and important
to derive limit distributions
of pseudovelocities of quantum walkers
on variety of plane lattices different
from the square lattices
and in the higher-dimensional lattices
\cite{CLXGKK05}.
For example, the quantum coin of the Grover walk
in the $D$-dimensional hyper-cubic lattice
is given by the $2D \times 2D$
orthogonal matrix
$A^{(D)}=(A^{(D)}_{jk})$
with the elements
\begin{equation}
A^{(D)}_{jk} = \Bigg\{
\begin{array}{ll}
1/D-1, \quad & \mbox{if $j=k$} \cr
1/D, \quad & \mbox{if $j \not=k$}.
\end{array}
\label{eqn:AD}
\end{equation}
It is also an interesting problem to relate
the present results to solutions of
the continuous-time quantum-walk models
on two-dimensional lattices \cite{MVB05}.

At the end of the present paper, 
we refer to the fact that
recent papers propose implementations of
not only one-dimensional
but also two-dimensional quantum walks
using optical equipments \cite{RS05,EMBL05},
ion-trap systems \cite{FOBH05},
and ultra-cold Rydberg atoms 
in optical lattices \cite{CREG06,MBA07}.
We hope that combinations of experiments
and theoretical works of quantum physics
will make significant contribution
to development of quantum informatics.

%%%%%%%%%%%%%%%%%%%%%%%%%%%%%%%%%%%%%%%%%%%%%%%%%%
\begin{acknowledgments}
M. K. would like to thank Norio Inui for useful comments
on the manuscript.
This work was partially supported by the Grant-in-Aid
for Scientific Research (C) (No. 17540363) of
Japan Society for the Promotion of Science.
\end{acknowledgments}

%%%%%%%%%%% APPENDICES %%%%%%%%%%%%%%%%%%%%%%%%%%%%%%%%
\appendix
%%%%%%%%%%%%%% Appendix A %%%%%%%%%%%%%%%%%%%%%%%%%%%%%
\section{On Integrals}
%%%%%%%%%%%%%%%%%%%%%%%%%%%%%%%%%%%%%%%%%%%%%%%%%%%%%%%

Consider the integral
$$
I=\int_{-\infty}^{\infty} dv_x 
\int_{-\infty}^{\infty} dv_y 
{\bf 1}_{\{v_x^2/p+v_y^2/q < 1\}}
\frac{1}{(v_x+v_y+1)(v_x-v_y+1)(v_x+v_y-1)(v_x-v_y-1)}
$$
with $p+q=1, p,q \geq 0$.
Let
\begin{equation}
v_x = \sqrt{p} r \frac{1}{2} \left(z+\frac{1}{z} \right),
\quad
v_y = \sqrt{q} r \frac{1}{2i} \left(z-\frac{1}{z} \right).
\label{eqn:changeA1}
\end{equation}
Then
$$
I=-2^4 i \sqrt{pq} \int_0^1 dr \frac{J(r)}{r^3}
$$
with a contour integral on a complex plane $\C$,
$$
J(r)=\oint_{C_0} dz f(z),
$$
where
\begin{equation}
f(z)=\frac{z^3}{(z+z_+)(z+z_-)(z-z_+)(z-z_-)
(z+\overline{z_+})(z+\overline{z_-})
(z-\overline{z_+})(z-\overline{z_-})}
\label{eqn:fz}
\end{equation}
with
$$
z_{\pm}=(\sqrt{p}+i \sqrt{q})
\frac{1}{r}(1 \pm \sqrt{1-r^2}).
$$
Here $C_0$ denotes the unit circle centered at the origin
on $\C$, $|z|=1$.
There are four simple poles at
$z=z_-, \overline{z_-}, -z_-$
and $- \overline{z_-}$ inside of the contour $C_0$
and the Cauchy residue theorem can be applied
(see, for example, Chapter 4 in \cite{AF03})
to obtain
$$
J(r) = 2 \pi i
\Big\{ {\rm Res}(f, z_-) + {\rm Res}(f, \overline{z_-})
+{\rm Res}(f, -z_-)+{\rm Res}(f, -\overline{z_-}) \Big\},
$$
where we see 
\begin{eqnarray}
{\rm Res}(f, z_-)
&=&(z-z_-)f(z) \Big|_{z=z_-} \nonumber\\
&=& \frac{r^4}{2^7 \sqrt{pq} \sqrt{1-r^2}}
\frac{(\sqrt{p}+i\sqrt{q}\sqrt{1-r^2})
(\sqrt{q}-i \sqrt{p} \sqrt{1-r^2})}
{(1-pr^2)(1-qr^2)}
\nonumber
\end{eqnarray}
and ${\rm Res}(f, -z_-)={\rm Res}(f, z_-)$,
${\rm Res}(f, -\overline{z_-})
={\rm Res}(f, \overline{z_-})
=\overline{{\rm Res}(f, z_-)}$.
We obtain
$$
J(r)=\frac{\pi i}{2^{4}}
\frac{r^4}{\sqrt{1-r^2}}
\left\{ \frac{1}{1-pr^2}
+\frac{1}{1-qr^2} \right\}.
$$
The integral formula
\begin{equation}
\int_0^1 dx \frac{x}{(1-a^2 x^2) \sqrt{1-x^2}}
=\frac{\arcsin a}{a \sqrt{1-a^2}},
\quad |a| < 1
\label{eqn:formula1}
\end{equation}
is useful and we arrive at the result
$$
I=\pi(\arcsin \sqrt{p}
+\arcsin \sqrt{q})=\frac{\pi^2}{2}.
$$
It implies that $\mu_p(v_x, v_y)$
given by (\ref{eqn:bunpu1}) is well-normalized;
$\int_{-\infty}^{\infty} dv_x 
\int_{-\infty}^{\infty} dv_y \mu_p(v_x, v_y)
=I \times 2/\pi^2=1$.

Similarly, we can also calculate the integrals
\begin{eqnarray}
I_x &=& \int_{-\infty}^{\infty} dv_x 
\int_{-\infty}^{\infty} dv_y 
{\bf 1}_{\{v_x^2/p+v_y^2/q < 1\}}
\frac{v_x^2}{(v_x+v_y+1)(v_x-v_y+1)(v_x+v_y-1)(v_x-v_y-1)},
\nonumber\\
I_y &=& \int_{-\infty}^{\infty} dv_x 
\int_{-\infty}^{\infty} dv_y 
{\bf 1}_{\{v_x^2/p+v_y^2/q < 1\}}
\frac{v_y^2}{(v_x+v_y+1)(v_x-v_y+1)(v_x+v_y-1)(v_x-v_y-1)}.
\nonumber
\end{eqnarray}
By the change of integral variables (\ref{eqn:changeA1}),
we have
$$
I_x = - 2^2 i p \sqrt{pq}
\int_{0}^{1} dr \frac{J_x(r)}{r},
\quad
I_y = 2^2 i q \sqrt{pq}
\int_{0}^{1} dr \frac{J_y(r)}{r}
$$
with
$$
J_x(r)=\oint_{C_0} dz f_x(z), \quad
J_y(r)=\oint_{C_0} dz f_y(z),
$$
where
$f_x(z)=(z+1/z)^2 f(z)$ and
$f_y(z)=(z-1/z)^2 f(z)$
with (\ref{eqn:fz}).
The Cauchy residue theorem gives
$$
J_x(r) = \frac{\pi i}{2^2} 
\frac{r^4}{(1-pr)\sqrt{1-r^2}}, \quad
J_y(r) = -\frac{\pi i}{2^2} 
\frac{r^4}{(1-qr)\sqrt{1-r^2}}.
$$
The integral formula (\ref{eqn:formula1})
and the fact $\int_0^1 dr r/\sqrt{1-r^2}=1$
lead to the results
\begin{eqnarray}
I_x &=& \pi(\arcsin \sqrt{p}-\sqrt{pq}),
\nonumber\\
I_y &=& \pi(\arcsin \sqrt{q}-\sqrt{pq}).
\nonumber
\end{eqnarray}
Since $K_x=I_x \times (2/\pi^2)$
and $K_y=I_y \times (2/\pi^2)$,
they give the expression (\ref{eqn:Delta1}).

It is interesting to see that the above calculation
of the integral $I$ gives the following identity,
\begin{equation}
\frac{1}{2}\int_{-\infty}^{\infty} dv_x 
\int_{-\infty}^{\infty} dv_y \mu_p(v_x, v_y)
=\int_{0}^{\infty} dr \, r \mu(r; \sqrt{p})
+\int_{0}^{\infty} dr \, r \mu(r; \sqrt{q}),
\label{eqn:identity1}
\end{equation}
where $\mu(x;a)$ is the Konno density-function
of one-dimensional quantum walk
\cite{Kon02,Kon05,KFK05,MKK07,SKKK08}
$$
\mu(x;a)=\frac{\sqrt{1-a^2}}{\pi
(1-x^2) \sqrt{a^2-x^2}}
{\bf 1}_{\{|x| < |a|\}}.
$$

%%%%%%%%%%%%%%%%%%%%%%%%%%%%%%%%%%%%%%%%%%%%%%%%%%%%
%%%%%%%%%%%%%%%%%%%%%%%%%%%%%%%%%%%%%%%%%%%%%%%%%%%%
% Create the reference section using BibTeX:
%\bibliography{basename of .bib file}

\end{document}